\documentstyle[11pt,newpasp,twoside,epsf]{article}

\def\etal{{et\thinspace al.}}                 

\def\edcomment#1{\iffalse\marginpar{\raggedright\sl#1\/}\else\relax\fi}
\reversemarginpar

\begin{document}

\title{Diagnostics of composite starburst+Seyfert 2 nuclei: Hints on
the starburst-AGN connection}

\author{Roberto Cid Fernandes}
\affil{Depto.\ de F\'{\i}sica, UFSC, Po Box 476, 88040-900,
Florian\'opolis, SC, Brazil}
\author{Tim Heckman}
\affil{Johns Hopkins University, Baltimore, MD, USA}
\author{Henrique Schmitt}
\affil{National Radio Astronomy Observatory, Socorro, NM, USA}
\author{Rosa Maria Gonz\'alez Delgado}
\affil{Instituto de Astrof\'{\i}sica de Andaluc\'{\i}a (CSIC) , Granada, Spain}
\author{Thaisa Storchi-Bergmann}
\affil{Instituto de F\'{\i}sica, UFRGS, Porto Alegre, RS, Brazil}

\begin{abstract}
We present a simple population synthesis scheme which recognizes
composite starburst+Seyfert 2 nuclei from a few easy-to-obtain optical
measurements. Composite systems seem to evolve towards less luminous
Seyfert 2's which do not harbor detectable circum-nuclear starbursts.
We encourage applications of this cheap diagnostic tool to large
samples of Seyfert 2's, as well as its extension to other activity
classes, in order to test and refine this evolutionary scenario.
\end{abstract}

\section{Introduction}

The nowadays popular expression ``starburst-AGN connection'' was to
our knowledge first coined by Tim Heckman in a 1991 conference paper
which dealt primarily with the then hot debate over the starburst
model for AGN of Roberto Terlevich and collaborators.  Since then,
several groups have gathered unambiguous evidence that vigorous
star-formation occurs in the inner few hundred pc of many AGNs (see
Cid Fernandes \etal\ 2001a and references therein), flooding AGN
papers with considerations about the effects of starbursts. This very
volume contains several new reports of nuclei exhibiting both
starburst and AGN properties (e.g.\ Colina \etal, Levenson \etal) and
of ways of diagnosing such {\it composite} systems (e.g.\ Kohno
\etal). While these discoveries undoubtedly strengthen the long held
suspicion that these two phenomena are intertwined in some fundamental
way, in all fairness, we still do not know what this connection
actually means! In a way, the situation was perhaps clearer 5--10
years ago, when the discussion was polarized in terms of the pros and
cons of the starburst model for AGN, a central theme of the La Palma
meeting back in 1993 (Tenorio-Tagle 1994). Now that the focus has
(ironically) drifted away from that debate (because of the
overwhelming evidence that super-massive black-holes do exist,
gathered since we last meet here in ``la isla bonita''), there is not
a well defined theoretical framework able to make good use of these
new data.  Now that we all recognize that the issue is not starburst
{\it or} black-hole, the fundamental questions are (1) what role do
starbursts play in defining the observed properties of AGN, and (2)
what is the physics linking starbursts and AGN?

The kind of starburst-AGN connection which would do more justice to
the term ``connection'' is one in which circum-nuclear star-formation
somehow controls the accretion rate and/or vice-versa, via
symbiotic/feedback processes, perhaps on the lines of the old models by
Perry \& Dyson (1985) and Norman \& Scoville (1988). A more trivial
and less causal connection would be one that links starbursts and AGNs
by their common eating habits. Both starbursts and black-holes live on
gas, so feeding the inner regions of galaxies (by the dynamical
processes discussed in this meeting) may well lead to a simple genetic
link between star-formation and nuclear activity, with either
phenomenon proceeding essentially unaware of the concomitant
occurrence of the other.  These two extreme alternatives, which
broadly outline ``nurture or nature'' perspectives on the
starburst-AGN connection, respectively, are presently viable.  By the
time we next gather in La Palma we will surely have a clearer
understanding of which of them is more relevant.

Observational clues on the nature of the starburst-AGN connection
require the careful study of systems in which both phenomena
co-exist.  A critical first step is hence to devise ways of
identifying such composite starburst+AGN nuclei.  A second step is to
characterize the basic parameters of both the starburst (e.g., its
star-formation rate) and the AGN (say, its accretion rate), preferably
for as large a number of systems as possible, so that one can address
issues such as evolutionary effects, correlations between the
starburst and AGN properties, the frequency of starbursts in AGNs and
the role of the host galaxy (see Storchi-Bergmann's contribution
elsewhere in this volume).  In these few pages we summarize some of
our results concerning the first step, i.e., the diagnosis of
compositeness. Our major ``publicity'' goal here is to convince the
reader that we found relatively cheap ways of identifying composite
systems and to encourage her/him to apply them to her/his data sets.
We predict that such applications will substantially enlarge the
current database of composite systems, thus providing plenty of raw
material for more detailed studies, necessary to further our
understanding of the starburst-AGN connection.

\section{How to tell a starbursting from a ``boring'' Seyfert 2}

AGNs are so complex and interesting by themselves that AGN-auts have
traditionally been reluctant to meddle in the business of stellar
populations. The sentence ``removing the starlight contribution'',
present in so many papers since the late 70's, epitomizes the
historical view of stellar populations as an annoying pollution of AGN
spectra. Most of the recent advances in the topic of the starburst-AGN
connection stem from works which break this barrier by daring to
dissect this ``pollution'', using it to characterize the stellar
content within the central kpc of active galaxies. This lead to an
outbreak of discoveries of starbursts around AGN, predominantly
Seyfert 2's, where the nuclear light-house is conveniently blocked
from view, facilitating the study of circum-nuclear stellar
populations. For instance, stellar wind lines in the UV, the WR bump
and/or high order Balmer absorption lines, all signatures of recent or
ongoing star-formation, were detected in 15 out of 35 Seyfert 2's in
our combined northern (Heckman \etal\ 1997; Gonzalez Delgado, Heckman,
\& Leitherer 2001) and southern (Cid Fernandes, Storchi-Bergmann, \&
Schmitt 1998; Storchi Bergmann, Cid Fernandes, \& Schmitt 1998;
Storchi-Bergmann \etal\ 2000) samples.  Identifying these
finger-prints of starbursts required high S/N data and years of hard
work. But {\it can we make it simpler?} In other words, can we use
these data to figure out a less expensive way of identifying composite
starburst+Seyfert 2 systems?

\begin{figure}
\plotfiddle{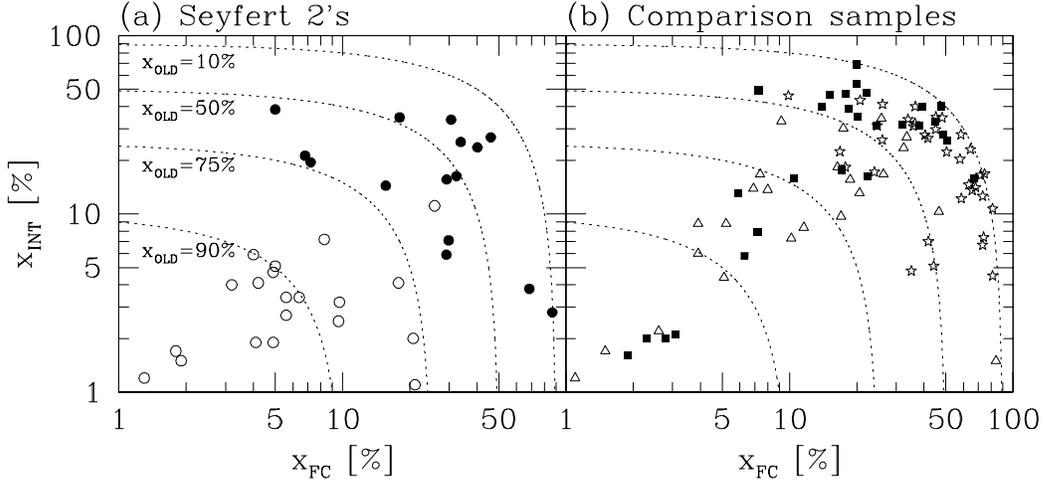}{6cm}{0}{70}{70}{-220}{-120}
\caption{Results of the synthesis analysis, condensed into a
($x_{FC}$,$x_{INT}$,$x_{OLD}$) representation. Dotted lines trace
lines of constant $x_{OLD}$. (a) Sources from our Seyfert 2 sample.
Filled circles indicate composite systems, while empty circles
indicate ``boring'' Seyfert 2's.  (b) Results for Starburst galaxies
(stars), Narrow Line AGN (empty triangles), some of which are
composites, and Mergers (filled squares).}
\end{figure}

The answer is {\it yes}! This answer was reached from an apparently
hopeless method. We simply feed the population synthesis code of Cid
Fernandes \etal\ (2001b; see also Schmitt, Storchi-Bergmann, \& Cid
Fernandes 1999 and Le\~ao \etal\ in this volume) with the equivalent
widths of Ca K, CN and the G-band absorption features, plus a couple
of near-UV colors ($F_{3660}/F_{4020}$ and $F_{4510}/F_{4020}$).  Very
few observables to solve an acknowledgedly difficult problem
(population synthesis), which, despite its ``long and venerable
history'' (Worthey 1994), has something of a ``bad reputation''
(Searle 1986).  Sure enough, we were not able to achieve a detailed
description of the stellar populations in terms of all age and
metallicity-related parameters in the code with so little
information. However, we achieved excellent results by reducing the
dimensionality of the problem to just 3 components: $x_{OLD}$, which
is the total fraction of light due to stars of $10^9$ yr or more,
$x_{INT}$, the fraction due to $10^8$ yr (post-starburst) populations,
and $x_{FC}$, which congregates all stellar generations of age $\le
10^7$ yr plus a power-law Featureless Continuum (FC), included to
account for scattered light.  While it is generally difficult to
decide whether young stars or a genuine AGN FC dominate this last
component (a historic problem which is still unsolved---see Storchi
Bergmann \etal\ 2000), in practice the strongest FC's are found in
composites, for which young stars clearly dominate $x_{FC}$. The
$x$-components, which we chose to normalize at the 4861 \AA\
continuum, must add up to 100\%, thus defining a plane in
($x_{OLD}$,$x_{INT}$,$x_{FC}$)-space, so in practice the method
provides a {\it bi-parametric} semi-empirical description of the data.

The results of this exercise are shown in Fig.~1a, where the synthetic
proportions are projected in the $x_{FC}$-$x_{INT}$ plane. Our 15
certified composite systems are plotted as filled circles, whereas
``boring'' Seyfert 2's (those where we have not detected signatures of
circum-nuclear starburst activity) are shown as empty circles.
Regardless of the actual meaning of the synthesis analysis, even the
most skeptical reader must agree that the method provides an excellent
empirical tool to separate composite from boring Seyfert 2's!  {\it
All} composites but {\it only one} ``pure'' Seyfert 2 (NGC 1068,
affected by its uniquely strong scattered light component) have
$x_{OLD} < 75\%$ and thus $x_{INT} + x_{FC} > 25\%$, a threshold which
can be equally well expressed by $W_{CaK} < 10$ \AA.  Also, {\it all}
$x_{FC}>30\%$ sources are composites.  A remarkable fact about this
diagram is that it does {\it not} use any of the information which was
used to classify systems as composite (UV lines, the WR bump and/or
high order Balmer lines). The synthesis picks out the composites
simply by their diluted metal absorption bands and blue colors. As
these data are relatively easy to obtain (all you need is a S/N $\ga
10$ optical spectrum), the method can be readily applied to large data
sets, offering a cheap way to find many more composites.  We have in
fact verified that the method works for other samples, as illustrated
in Fig.~1b (see Cid Fernandes \etal\ 2001a for details and for other
handy properties of the synthesis parameters).

\section{Evolution: Do composites end up as ``boring'' Seyfert 2's?}

Another remarkable fact about Fig.~1 is that our seemingly crude
synthesis yields a good description of the {\it evolutionary status}
of the circum-nuclear starbursts in composites.  The youngest
composites (those which exhibit O and/or WR star features, such as
Mrk 477) are located in the bottom-right part of the plot (large
$x_{FC}/x_{INT}$) whereas systems in a post-starburst phase (with
pronounced high order Balmer absorption lines, such as ESO 362-G8)
populate the top-left region, and nuclei which exhibit both
characteristics (e.g., NGC 5135, NGC 7130) are located in between
(top-right). The bottom-line here is that, more than a necessary evil,
betting on a simple population synthesis analysis of Seyfert 2's was
not a bad idea after all!

The evolutionary path of a starburst in Fig.~1 would roughly follow a
$x_{FC} \rightarrow x_{INT} \rightarrow x_{OLD}$ sequence as the burst
ages and fades, converging to bottom-left of the plot, i.e., the
region populated by ``boring'' Seyfert 2's. This naturally suggests
that ``boring'' Seyfert 2's are the end point of composites!  It must
be stressed, however, that (at least some) ``boring'' Seyfert 2's may
well have faint but not necessarily old starbursts, which are
naturally harder to detect against the dominant background of old
bulge stars (quantified by their large $x_{OLD}$). In fact, we came to
recognize that we know much less about the ``boring'' Seyfert 2's than
about the composite ones, which, somewhat paradoxically, makes
``boring'' Seyfert 2's more interesting! Despite such caveats, such an
evolutionary scenario seems compelling.

\section{Composites are more luminous than ``boring'' Seyfert 2's}

\begin{figure}
\plotfiddle{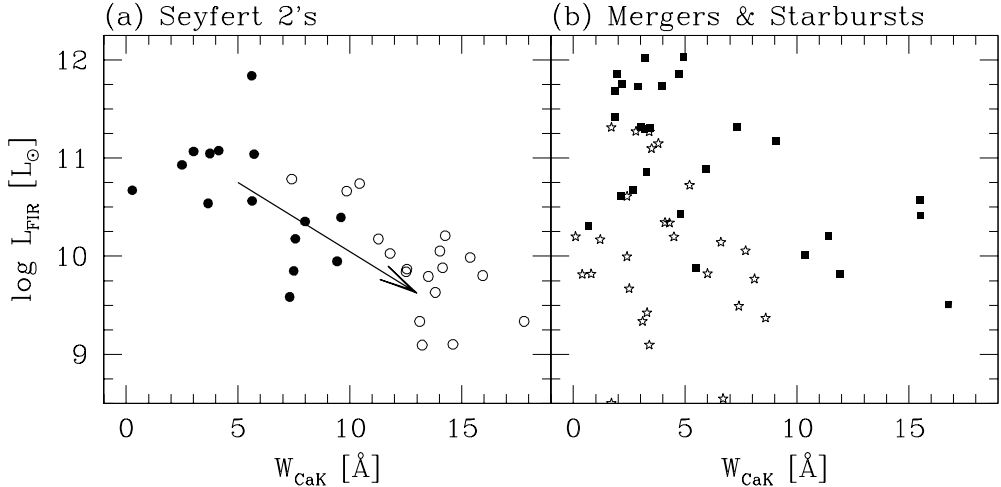}{6cm}{0}{70}{70}{-220}{-120}
\caption{Far IR (60 + 100 $\mu$) luminosity against $W_{CaK}$, an
empirical diagnostic of compositeness, for (a) the Seyfert 2 sample,
and (b) Starburst and Merger galaxies. Symbols as in Fig.~1.  A
qualitative evolutionary arrow is sketched. Comparing the two panels
we see that as a whole our Seyfert 2 sample is rather comparable to
merger systems.}
\end{figure}

Among the many interesting systematic differences we found between
composite and ``boring'' Seyfert 2's is that the former are
significantly {\it more luminous} than the latter. This is illustrated
in Fig.~2, where we see that composites are typically $\sim 5$ times
more luminous in terms of their far IR emission than ``boring''
Seyfert 2's. A more dramatic way of putting it is that above $L_{FIR}
= 2 \times 10^{10}$ L$_\odot$, $\sim 80\%$ of Seyfert 2's in our
sample are composites. Luminous Seyfert 2's may therefore owe much of
their luminosity output to circum-nuclear star-formation. Composites
are also more luminous in the optical continuum and emission lines, a
difference which is only exacerbated when reddening corrections are
applied, since composites are also more optically extincted than
``boring'' Seyfert 2's, consistent with the X-ray analysis of
Levenson, Heckman \& Weaver (2001).  Intuitively, one expects younger
things to be more luminous, so this tendency fits well in the
evolutionary scenario outlined by the (luminosity-independent)
synthesis analysis. We also find that ``boring'' Seyfert 2's are
weaker both in indices sensitive to starburst ($L_{FIR}$,
$L_{H\beta}$) and AGN ($L_{[\ion{O}{III}]}$) activity, indicating that
more luminous AGN host correspondingly more luminous circum-nuclear
starbursts! A more refined analysis may eventually translate this
finding into something like a ``star formation rate $\propto$
accretion rate'' relation.

\section{Final remarks: Starbursts are here to stay}

In our full comparative study of the properties of composite and
``boring'' Seyfert 2's (which includes the analysis of reference
samples of Starburst, active, merging and normal galaxies), we further
identify the effects of circum-nuclear starbursts upon emission line
equivalent widths, gas excitation indices, line profiles, near-UV
surface brightness and central morphology.  Whilst we still lack a
clear understanding of whether circum-nuclear starbursts are indeed an
integral part of the AGN phenomenon, there remains no doubt that they
exist and play a major role in defining observable properties
traditionally attributed solely to AGN.  Like it or not, theorists
have to deal with this fact.  For instance, we find that typically
50\% (and up to 80\%) of the nuclear H$\beta$ luminosity of composites
is powered by massive stars!  At the very least, this calls for a
revision of classical photoionization models, which must mix a
starburst and a harder ionizing source when modeling nuclear emission
lines in Seyfert 2's. Less starburst-phobic readers will see these
results as further signals of a truly physical connection between
star-formation and nuclear activity.  Establishing how fundamental
this connection is will require the extension of diagnostics such as
the ones presented here to larger samples of Seyfert 2's in order to
derive statistically robust results, as well as a more panoramic view
of starbursts in AGNs as a whole, from LINERs to QSOs.
Slowly, but surely, we are putting together the pieces of this puzzle.

\end{document}